\documentclass[12pt,preprint]{aastex}
\usepackage{natbib}
\usepackage{amssymb,amsmath} 

\shorttitle{The measurement of the speed of GW}
\shortauthors{D. Shuang}

\begin{document}

\title{On the measurement of the speed of gravitational wave}

\author{Shuang Du \altaffilmark{1,2}}
 \altaffiltext{1}{Department of Physics and Astronomy, Sun Yat-Sen University, Zhuhai 519082, China; dushuang@pku.edu.cn}
 \altaffiltext{2}{School of Physics, Peking University, Beijing 100871, China}

\begin{abstract}
Strict measurement of the speed of gravitational wave (GW) is very important for fundamental physics.
In this paper, taking cosmological effect into account, we derive a more precise formula for calculating the speed of GW based on GW 170817-like events.
We find that in the case of high redshift, the usual luminosity distance needs to be replaced by co-moving distance.
We consider some possible electromagnetic signals which can be significantly reduce the intrinsic uncertainty between the
emission times of the gravitational and electromagnetic signals. Hopefully, the measurement accuracy will be improved to $\sim 10^{-18}\;\rm m\; s^{-1}$.
\end{abstract}


\section{Introduction}\label{sec1}
Strict measurement of the speed of gravitational waves (GWs) is very important for fundamental physics.
Exact value of the speed of GW can be used to test the theories of gravity (see Will 2014 for review) and the continuity of spacetime (Kosteleck{\'y} \& Samuel 1989, Amelino-Camelia et al. 1998).
Gravitational wave-electromagnetic wave (GW-EW) association (e.g., GW 170817/GRB 170817A/kilonova
AT2017gfo association; Coulter et al. 2017; Arcavi et al. 2017; Abbott et al. 2017a) provides an opportunity to gain
insight into these problems. Besides, the complementarity of the information derived from the GWs and EWs
of the same source shows a strong potential in solving some long-standing problems about the GW-EW sources
themselves (Hayama et al. 2016; Du et al. 2018).

By comparing the arrival times of GWs and gamma-ray photons from the double neutron star (NS) merger,
the speed of GWs is constrained as (Abbott et al. 2017a)
\begin{equation}\label{EQ}
-3\times 10^{-15}\leq \frac{\Delta v_{\rm GW}}{c}\leq +7\times 10^{-16},
\end{equation}
where $\Delta v_{\rm GW}$ is the difference between the speed of GWs and the speed of light.
But it is a pity that the intrinsic difference in emission time between the GWs and gamma-ray photons is uncertain
and depends on the unknown jet physics of gamma-ray bursts (GRBs)\footnote{The current situation is that some
authors constrain GRB physics by using GWs travel at the speed of light, while others use GRB physics to constrain the speed of GWs. }.
One the other hand, before measuring the speed of GWs,
we must assume that they travel at different speed than the speed of light, such that the path of GWs may be
timelike geodesic (Fan et al. 2017)\footnote{Fan et al. (2017) also propose a method to measure the speed of GWs.
But the accuracy is relatively low $\sim 10^{-10}\; \rm m\; s^{-1}$.}
However the path of photons is null geodesic. That's to say, the velocity in cosmology should be specified.

Given the importance of the speed of GWs to the fundamental physics, we should take this issue more seriously.
It¡¯s also quite important from the perspective of surveying that, as a basic physical quantity, the speed of GWs needs to be measured as accurately as possible.
Therefore, at first, we need some new GW-EW sources which only has a small intrinsic uncertainty between the
emission times of the gravitational and electromagnetic signals .

We have known that double NS mergers would lead to short GRBs. The interaction between the magnetospheres of two NSs before the last merger
should also generate a electromagnetic signal. According to the unipolar inductor model of close binaries (Piddington \&
Drake 1968; Goldreich \& Lynden-Bell 1969), the motion of the weakly magnetized companion relative to the magnetic field of the strongly magnetized primary will induce
a electromotive force, and thus electromagnetic radiation can be generated by the accelerated charged particles.
Piro (2012), Lai (2012), and Wang et al. (2016) studied
this electromagnetic radiation in detail under the scenario that a NS binary in the later period of inspiral.

On the other hand, for two stars that are close to each other, if their magnetic fields have different orientations,
there should be also a electromagnetic radiation due to magnetic reconnection mechanism (Uchida 1986). Here,
this close binary system is also restricted to the NS binary (also see Wang et al. 2018).
We will focus on the detectability of such electromagnetic signal.
This paper is organized as follows. We derive a more precise formula to calculate the speed of GWs by taking cosmological effect into account in Section 2.
We investigate the properties of the electromagnetic radiation in the last inspiral phase of the NS binary system according
to the magnetic reconnection mechanism in Section 3. We use these GW-EW associations to constrain
the speed of GWs in Section 4. Section 5 is the discussion.

\section{The formular to calculate the GW speed}\label{sec.2}
To measure the speed of GWs, one needs to define the speed which is measurable and isotropic first.
Under the Friedman-Robertson-Walker metric
\begin{eqnarray} \label{Eq9}
ds^{2}=-c^{2}dt^{2}+a(t)^{2}\left(dr^{2}+r^{2}d\theta^{2}+r^{2}\sin^{2}\theta d\phi^{2}\right),
\end{eqnarray}
the definition of velocity with practical physical meaning is proper velocity $v_{\rm p}$.
Taking the line of sight as the radial direction, there is
\begin{eqnarray} \label{Eq10}
v_{\rm p}=\frac{a(t)dr}{dt}.
\end{eqnarray}
For a photon, $ds=0$, it is easy to verify that the speed of light is $c$ under this definition.
Since the co-moving coordinate $r$ of the GW-EW source is an invariant, according to equation (\ref{Eq10}), the speed of GWs $v_{\rm GW}$ and the speed of light satisfy
\begin{eqnarray} \label{Eq11}
\int_{t_{\rm e2}}^{t_{\rm mer}} v_{\rm GW}\frac{dt}{a(t)}=\int_{t_{\rm e1}}^{t_{\rm tou}}c\frac{d{t}'}{a({t}')},
\end{eqnarray}
where $t_{\rm e1}$ and $t_{\rm tou}$ is the emission time and arrival time of EWs, and $t_{\rm e2}$ and $t_{\rm mer}$ is the the emission time and arrival time of GWs, respectively.
Equation (\ref{Eq11}) also can be rewritten as
\begin{eqnarray}\label{Eq12}
&&\int _{t_{\rm e2}}^{t_{\rm e1}}v_{\rm GW}\frac{dt^{\prime}}{a(t^{\prime})}+\int _{t_{\rm tou}}^{t_{\rm mer}}v_{\rm GW}\frac{dt^{\prime\prime}}{a(t^{\prime\prime})}\nonumber\\
&&=\int _{t_{\rm e1}}^{t_{\rm tou}}(c-v_{\rm GW})\frac{dt}{a(t)}.
\end{eqnarray}
Now the difficulty is that $t_{\rm e1}$ and $t_{\rm e2}$  are uncertain, as well as $\Delta t=t_{\rm e2}-t_{\rm e1}$.
This is exactly the difficulty confronted Abbott et al. (2017a).
We assume that there are some sources (see Section 3) whose $\Delta t$ is much smaller than the travel times of GWs and EWs, i.e., $t_{\rm mer}-t_{\rm e2}$ and $t_{\rm tou}-t_{\rm e1}$.
On the other hand, observations (Will 2014; Abbott et al. 2017a) show that $t_{\rm mer}-t_{\rm tou}$
should be also a small quantity when compared to the travel times of GWs and EWs.
Therefore, equation (\ref{Eq12}) is approximatively given by
\begin{eqnarray}\label{Eq14}
\frac{v_{\rm GW}}{1+z_{\rm e}}(t_{\rm e1}-t_{\rm e2})+v_{\rm GW}(t_{\rm mer}-t_{\rm tou})\nonumber\\
=(c-v_{\rm GW})\int_{0}^{z_{\rm e}}\frac{dz}{H(z)},
\end{eqnarray}
where $z_{\rm e}$ is the redshift of the NS binary, and $H(z)$ is the Hubble parameter.
Since co-moving distance is
\begin{eqnarray}\label{Eq15}
D_{\rm c}=c\int_{0}^{z_{\rm e}}\frac{dz}{H(z)},
\end{eqnarray}
following from equation (\ref{Eq14}), one has
\begin{eqnarray}\label{Eq16}
v_{\rm GW}&=&\frac{c}{1+\frac{c}{D_{\rm c}}\left[(t_{\rm mer}-t_{\rm tou})-\frac{t_{\rm e2}-t_{\rm e1}}{1+z_{e}}\right]}\nonumber\\
&\approx&c\left \{ 1-\frac{c}{D_{\rm c}}\left[(t_{\rm mer}-t_{\rm tou})-\frac{\Delta t}{1+z_{e}}\right] \right \}.
\end{eqnarray}
Therefore, equation (\ref{EQ}) is modified to
\begin{eqnarray}
\frac{c-v_{\rm GW}}{c}=\frac{1}{D_{\rm c}}\left[(t_{\rm mer}-t_{\rm tou})-\frac{\Delta t}{1+z_{e}}\right].
\end{eqnarray}
Assuming that the measurement error of arrival times is much less than the uncertainty of $\Delta t$ (i.e., $\delta t$ in equation \ref{Eq17}),
according to equation (\ref{Eq16}), the uncertainty of the speed of GWs $\delta v_{\rm GW}$ is
\begin{eqnarray}\label{Eq17}
\frac{\delta v_{\rm GW}}{c}=\frac{1}{c}\frac{\partial v_{\rm GW}}{\partial \Delta t}\delta t=\frac{c}{D_{\rm c}}\frac{\delta t}{1+z_{\rm e}}.
\end{eqnarray}
It is clear that one should reduce the value of $\delta t$ to improve the measurement accuracy.

\section{Electromagnetic precursor from double NS merger }\label{sec.3}
\begin{figure*}
\centering
  \includegraphics[width=0.33\textwidth]{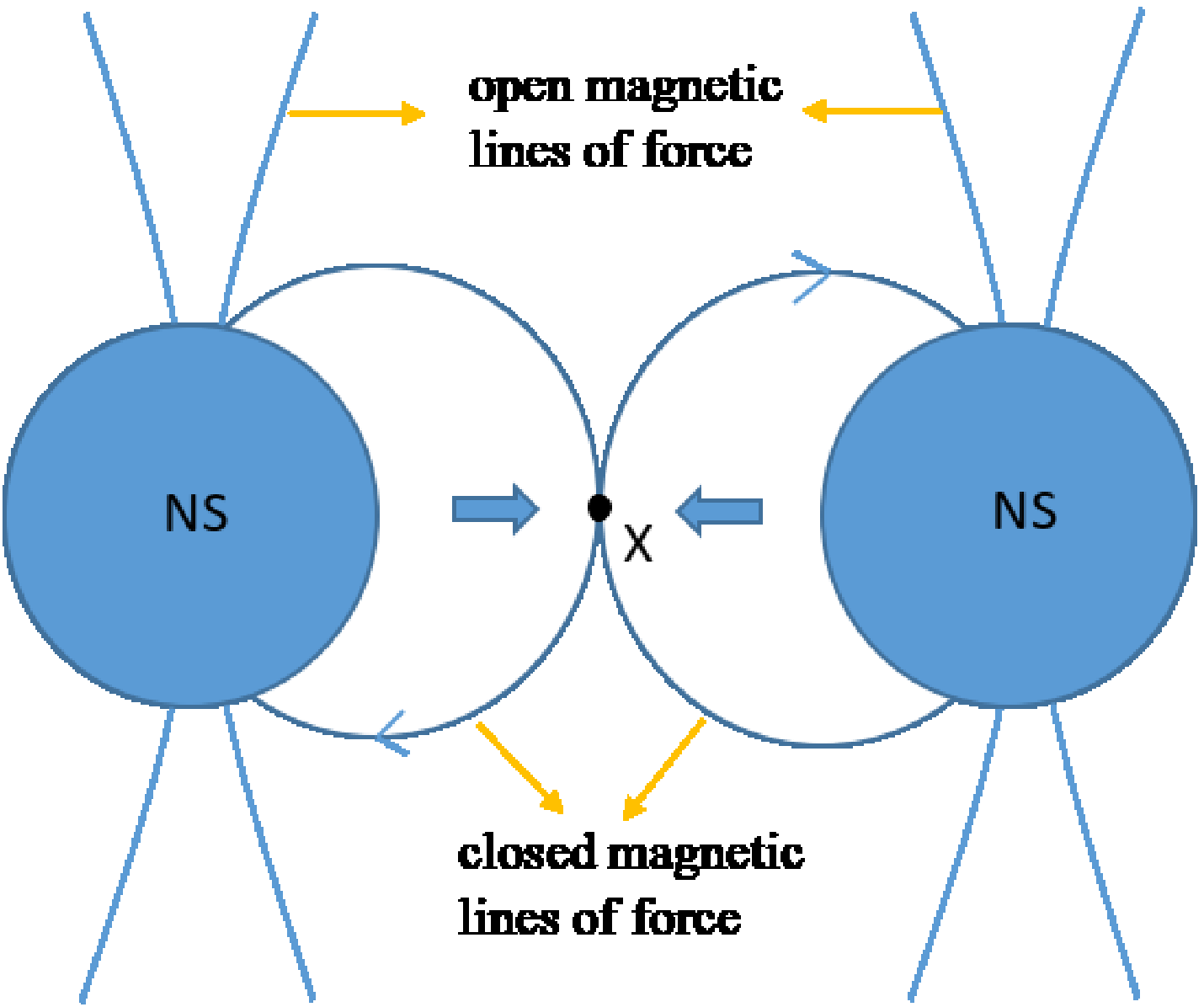}
  \includegraphics[width=0.33\textwidth]{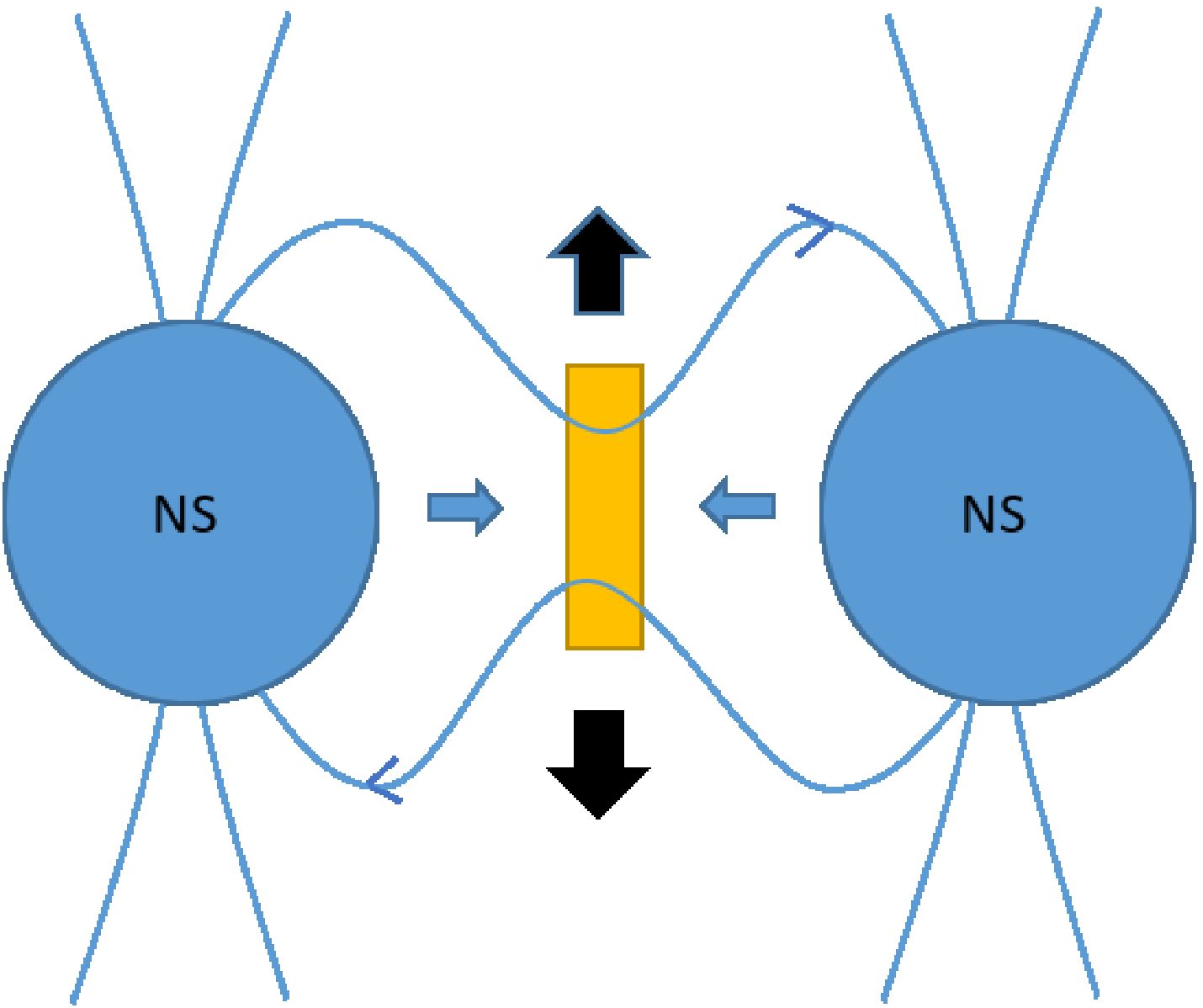}
  \caption{A schematic diagram to the magnetic reconnection of NS binary system. Left panel:
  Carried by the inflows (blue solid arrows), the magnetic lines of force with opposite directions come into contact at $\rm X$ point.
Right panel: Magnetic field lines break and reconnect in the dissipation region (yellow region), which then produces outflows (black solid arrows).
  \label{fig.1}}
\end{figure*}

As shown in Fig. \ref{fig.1}, the magnetic dipole fields of the two NSs are anti-parallel.
The orbital separation between the two NSs is $a$.
For simplicity, the radii of the two NSs $R_{\ast}$ and the dipole magnetic field
strength on the two NSs surfaces $B_{\ast}$ are both assumed to be similar.
According to the mechanism of magnetic reconnection, when the two NSs carry the plasma
located in their magnetospheres to close to each other (inflows, see the blue solid
arrows), the collision between these two inflows
in the dissipation region (yellow region) may lead to the disruption and reconnection of magnetic lines
of force. At the same time, magnetic energy will translate into kinetic energy and thermal
energy of the plasma.

The dissipation rate of magnetic energy $\dot{E}_{\rm B}$ depends on the velocity of inflow
$v_{\rm in}$. But since the magnetosphere environment is uncertain, $v_{\rm in}$ is an unknown parameter.
Second best, if only for rough estimation, one can use the characteristic parameters of the NS binary
system to estimate $\dot{E}_{\rm B}$ through dimension analysis. Assuming a steady-state magnetic reconnection
that magnetic flux dissipated in the dissipation region is balanced by the inflows, one simply has
\begin{equation} \label{Eq1}
\dot{E}_{\rm B}\sim \frac{1}{2u_{0}}\left [ B_{\ast}\left ( \frac{a}{2R_{\ast}} \right )^{-3} \right ]^{2} \left (\dot{a}R_{\ast}^{2} \right ),
\end{equation}
where $u_{0}$ is the permeability of vacuum, and $v_{\rm in}\sim \dot{a}$ is adopted.
The time evolution of $a$ is (Wang et al. 2016)
\begin{equation} \label{Eq2}
a=20\left ( 1-1695t\right )^{1/4}\;\rm km,
\end{equation}
where $a = 20\; \rm km$ at $t = 0$ is set when the surfaces of the two NSs just touch with each other.
Combining equations (\ref{Eq1}) and (\ref{Eq2}), one has
\begin{equation}\label{Eq3}
\dot{E}_{\rm B}\sim 10^{44}(1-1695t)^{-9/4}\left ( \frac{B_{\ast}}{10^{12}\;\rm Gs} \right )^{2}\;\rm erg\cdot s^{-1},
\end{equation}
here $R_{\ast}=10\;\rm km$ is adopt \footnote{There is a difference in the exponential compared to the equation (2) of Wang et al. (2018), since we use a different method.
But during the last inspiral (which is what we are interested in), the values of $\dot{E}_{\rm B}$ predicted by the manuscript and Wang et al. (2018) are coincident.}.
The dissipation of magnetic energy will accelerate electrons (electrons and positrons are collectively referred to electrons). The energy of electrons can be
converted to electromagnetic radiation through, e.g., synchrotron radiation.
The peak frequency of synchrotron radiation is
\begin{equation}\label{Eq4}
\nu_{\rm m}\approx 1.3\times10^{18}\left (\frac{\Gamma }{10^{2}}\right )^{2}\left ( \frac{B}{10^{8}\;\rm Gs }\right )\; \rm Hz,
\end{equation}
where $\Gamma$ is the Lorentz factor of electrons, and $B$ is the magnetic field strength at the region where the electrons begin to radiate.
Empirically, the intense radiations of NSs in the X-ray band are generally believed to be related to the dissipation of magnetic field, such as
soft gamma-ray repeaters (Katz 1982; Thompson \& Duncan 1995), anomalous X-ray pulsars (Thompson \& Duncan 1996), and X-ray flares in GRBs (Dai et al. 2006).
So we take the efficiency of the transformation of magnetic energy into keV emission as $\eta=0.1$
(see Appendix for a qualitative discussion; see Figure 3 in Wang et al. (2018) for numerical calculation).
Based on the flux sensitivity of \emph{Fermi} BAT that $F_{\rm min}=0.07\; \rm photons\; cm^{-2}\; s^{-1}$, and following the method of McWilliams \& Levin (2011),
this keV radiation can be detected at
\begin{eqnarray} \label{Eq5}
D_{\rm L}&=&\sqrt{\frac{\eta \dot{E}_{\rm B}}{\Delta\Omega F_{\rm min}}}\nonumber\\
&\approx& 120\left (\frac{\eta \dot{E}_{\rm B}}{10^{43}\;\rm erg\cdot s^{-1}}\right )\left(\frac{\Delta\Omega}{4\times 10^{4}\;\rm deg^{2}}\right)^{-1/2} \;\rm Mpc,
\end{eqnarray}
where $\Delta\Omega$ is the beaming angle of the keV emission.

In addition to synchrotron radiation, curvature radiation can also cool the electrons.
The cooling timescale due to curvature radiation is
\begin{equation} \label{Eq6}
t_{\rm cur}\approx 2\times 10^{8}\left (\frac{\rho}{10\;\rm km} \right )^{2}\left (\frac{\Gamma}{10^{2}} \right )^{-3}\;\rm s,
\end{equation}
where $\rho$ is the curvature radius of magnetic lines of force.
For comparison, the synchrotron-radiation lifetime of electrons is
\begin{equation} \label{Eq7}
t_{\rm syn}\approx 5\times 10^{-10}\left (\frac{\Gamma}{10^{2}} \right )^{-1}\left (\frac{B}{10^{8}\;\rm Gs} \right )^{-2} \;\rm s.
\end{equation}
It is clear that the cooling of electrons is dominated by the synchrotron radiation.

However, it is worth noting that there may be a giant radio pulse due to the curvature radiation.
According to the Ruderman-Sutherland model (Ruderman \& Sutherland 1975),
the electrons with $\Gamma\sim 10^{2}-10^{3}$ will make contribution to radio emission
through coherent curvature radiation in the region where the magnetic field lines are open (see Fig. \ref{fig.1}).
The location of the radio emission is $\rho_{\rm c}\sim 10^{8}\;\rm cm-10^{9}\;\rm cm$ away from the NS surface.
If the opening angle of the outflow satisfies $\theta>\rho_{\rm c}/R_{\ast}>10^{-3}\;\rm rad$,
some of the accelerated electrons in the outflow will enter this region and generate radio emission.
Following the method of Totani (2013), the flux in observed frequency $\nu_{\rm obs}$ is
\begin{eqnarray}\label{Eq8}
F_{\rm \nu }&=&\frac{1}{\nu _{\rm obs}}\frac{\epsilon _{\rm r}\left | \dot{E} \right |}{4\mathrm{\pi}D_{\rm L}^{2}} \nonumber\\
&\approx&0.2\left (\frac{\epsilon _{\rm r}}{10^{-3}}  \right )\left (\frac{\nu _{\rm obs}}{1.4\;\rm GHz}  \right )^{-1}\left  (\frac{D_{\rm L}}{300\;\rm Mpc}  \right )^{-2}\nonumber\\
&&\times \left(\frac{\dot{E}_{\rm B}}{10^{44}\;\rm erg\cdot s^{-1}}  \right )\;\rm Jy,
\end{eqnarray}
where $\epsilon _{\rm r}$ is the efficiency of converting magnetic energy into radio emission.
So this pulse may also be detected at cosmological distance.

\section{Constraining the speed of GWs}\label{sec.4}
Note that when the surfaces of the two NSs contact, the gap between the two NSs disappears and the magnetic reconnection occurs
in the two stars. The radiation of the accelerated electrons after this contact will be blocked by the dense star matter.
And the electrons accelerated before the collision quickly lose energy due to the keV emission (see equation \ref{Eq7}).
Therefore, the luminosity of keV emission will sharply decay, which will appear as a peak in the light curve.
But the same cannot be true for the gravitational radiation, the luminosity of gravitational radiation will keep increasing until the ringdown of merger remnant begins
(which is similar to the Figure 2 in McWilliams \& Levin 2011).
Except for X-ray emission, the radio emissions predicted here and by Totani (2013) and Wang et al. (2016) are
also very close to the last merger. Compared to GW-GRB associations, these GW-EW associations have
much less uncertainties in the emission times of gravitational and electromagnetic signals.

From now on, we use the original $t_{\rm e1}$ and $t_{\rm e2}$ to represent the emission times of peak GW and EW signals.
Then $\Delta t=t_{\rm e2}-t_{\rm e1}$ is roughly the duration time of NS binary merger ($R_{\ast}/\dot{a}(t=0)\sim 10\;\rm ms$).
The uncertainty $\delta t$ should not be larger than the duration time of NS binary merger itself, i.e., $\max(\delta t)\sim \mathcal{O}(10\;\rm ms)$.
Conservatively, we assume these GW and EW signals are detected at low redshifts, i.e., $z_{\rm e}\ll 1$,
equation (\ref{Eq17}) finally can be reduced to
\begin{eqnarray}\label{Eq18}
\frac{\delta v_{\rm GW}}{c}\sim 1\times 10^{-18}\left ( \frac{D_{\rm L}}{100\;\rm Mpc} \right )^{-1}\left ( \frac{\delta t}{10\;\rm ms} \right ).
\end{eqnarray}

\section{Discussion}
In this paper, a more precise formula to calculate the speed of GWs is derived.
We show the possible signatures of the electromagnetic precursors before the last mergers of NS binaries,
and discussed their applications in constraining the speed of GWs.
We aim to provide a new strategy to strictly measure the speed of GWs.
The event rate of NS binary merger is $1540_{-1220}^{+3200}\;\rm Gpc^{-3}yr^{-1}$ (Abbott et al. 2017b).
There are enough sources to test our method in the range of $\sim 100\;\rm Mpc$.
In the future, space-GW detectors (e.g., LISA (Stroeer \& Vecchio 2006) and Tian Qin (Luo et al. 2016))
are able to search for NS binary systems and provide early forecasts for EW and high-frequency GW detectors,
such that the follow-up observations can determine $t_{\rm tou}$ and $t_{\rm mer}$.

In the above argument, the measuring errors of $t_{\rm tou}$ and $t_{\rm mer}$ are assumed to be secondary.
This is not always the case. If the luminosity is low and the sensitivity of the instruments is not high enough,
the measurement is not necessarily accurate. But to say the least,
there is always an opportunity to improve the technology to achieve higher time resolution.
Furthermore, when $t_{\rm tou}$ and $t_{\rm mer}$ are measured accurately, there is another intriguing implication.
Assuming that $v_{\rm GW}=c$, one has $t_{\rm mer}-t_{\rm tou}=t_{\rm e2}-t_{\rm e1}$ (At present, LIGO can not determine $t_{\rm mer}$,
a possible way can be found in van Putten \& Della Valle (2019).).
The value of $t_{\rm e2}-t_{\rm e1}$ is only depend on the radii of the two NSs under tidal deformation,
such that it may be useful to constrain the equation of state of NSs.

\section{Acknowledgement}
We thank Fang-Kun Peng for the help in writing this paper.
We thank Xiao-Dong Li, Yi-Jung Yang and Yi-Ming Hu for useful discussions.

\begin{appendix}
\section{Appendix} \label{app}
Near the NS, the magnetic field intensity is still strong despite the residual magnetic field after the magnetic reconnection.
Because the magnetic multipole field ($B\sim 10^{-4}B_{\ast}$ is adopted) can be retained for the random direction.
The acceleration of electrons will be suppressed by synchrotron radiation, which makes $\Gamma$ not very high.
Even high-energy $\gamma$-ray photons ($>\sim 1\;\rm MeV$) can be produced,
the cascade processes $e+B\rightarrow \gamma+ B$ and $\gamma+B\rightarrow e+e^{+}$ would convert them into electron-position pairs.
Specially, if the energy of $\gamma$-ray photons beyond $\sim 10^{2} \; \rm MeV$, the Lorentz factor of secondary electrons will be $\Gamma \sim 10^{2}$,
and the keV emission will appear (see equation \ref{Eq4}). If the energy of $\gamma$-ray photons is only several $\rm MeV$,
the emission of secondary electrons will be weak.
So it is expected that considerable magnetic energy can be transformed into keV photons through synchrotron radiation.
\end{appendix}

\end{document}